# Plasmonic modes in periodic metal nanoparticle chains: A direct dynamic eigenmode analysis

Kin Hung Fung and C. T. Chan[*]

*Department of Physics, The Hong Kong University of Science and Technology,*
*Clear Water Bay, Hong Kong, China*

We demonstrate an efficient eigen-decomposition method for analyzing the guided modes in metal nanoparticle chains. The proposed method has the advantage of showing the dispersion relation and mode quality simultaneously. It can also separate the material and geometrical properties, so its efficiency does not depend on the complexity of the material polarizability. The method is applied to analyze the guided modes of a single and a pair of metal chains. The more accurate dynamic dipole polarizability typically gives a red-shift compared with the results obtained with the broadly used quasistatic dipole polarizability with radiation correction.

There are recent interests of using metal nanoparticle (MNP) chains to function as a designable subwavelength waveguide to transport optical signals.[1,2] Wave propagation along MNP chains is mediated by coupled plasmonic resonances. Such plasmonic waveguides may serve as the building blocks for making nanoscale optical devices.[3,4] However, energy loss due to radiation and absorption in MNPs can be serious, and thus the design of low loss MNP waveguides become a priority concern.[5-11]

To investigate the waveguide modes of periodic MNP chains, some authors calculated the dispersion relations.[2,12-17] In the presence of dissipation, the guided modes have a finite life-time, and cannot be described completely by a simple dispersion relation between the real $\omega$ and real $k$. It is customary to allow either one of $\omega$ and $k$ to be a complex number. Operationally, this requires root searching of a complex function in the complex number plane. Such kind of root searching is time consuming, especially when full dynamic effect is considered. Some authors, therefore, only calculate the dispersion relation within quasistatic approximation.[2,14] To include the retardation effect associated with radiation, other authors use the quasistatic dipole approximation (DA) with radiation correction.[12,13,15,16] However, such an approximation is not accurate for particle size greater than 50 nm when absorption loss is not negligible.[18] Here, we demonstrate an eigen decomposition (ED) method[19,20] with a more accurate dynamic DA[21,22] to account for material properties. Such method does not require a root searching in the complex plane. The method consider both $\omega$ and $k$ on the real line and, at the same time, shows effectively the qualities of the guided modes.

We begin by considering an infinite periodic chain of MNPs along the $z$-direction, with particle diameter $d = 50$ nm and center-to-center distance of adjacent particles $a = 75$ nm. The permitivity of the MNP is $\varepsilon(\omega) = 1 - \omega_p^2/\omega(\omega + i\gamma)$, where $\omega_p = 6.18$ $eV$ and $\gamma$ is the electron scattering rate. Each particle is represented by an electric dipole $\mathbf{p}_m$ with polarizability $\alpha(\omega)$, where $m$ is the particle index. For a given external field $\mathbf{E}_m^{drive}$, the dipole moment of the m$^{th}$ particle (at a frequency $\omega$) is

$$\mathbf{p}_m = \alpha(\omega)\left[\sum_{m'\neq m}\vec{\mathbf{g}}(\mathbf{r}_m - \mathbf{r}_{m'})\cdot\mathbf{p}_{m'} + \mathbf{E}_m^{drive}\right], \quad (1)$$

where $\vec{\mathbf{g}}(\mathbf{r})$ is the dyadic Green's function. Many studies employ the quasistatic DA, which uses $\alpha(\omega) = d^3(\varepsilon(\omega)-1)/(\varepsilon(\omega)+2)$ with the radiation correction $\alpha^{-1}(\omega) \to \alpha^{-1}(\omega) - 2i\omega^3/(3c^3)$, where $c$ is the speed of light. For particle size $\approx 50$ nm with non-negligible absorption loss, a more accurate dynamic polarizability has to be used:[21,22]

$$\alpha(\omega) = i\frac{3}{2}\frac{c^3}{\omega^3}a_1(\omega), \quad (2)$$

where $a_1(\omega)$ is the electric dipolar term of the Mie coefficients[23]. The dipole approximation that uses the polarizability defined in Equation (2) is sometimes called dynamic DA (or exact dipole approximation[22]).

Eq. (1) can be written formally as

$$\mathbf{M}(\omega)|\mathbf{p}\rangle = |\mathbf{E}^{drive}\rangle, \quad (3)$$

where $|\mathbf{p}\rangle \doteq (...,\mathbf{p}_{m-1},\mathbf{p}_m,\mathbf{p}_{m+1},...)$ and $|\mathbf{E}^{drive}\rangle \doteq (...,\mathbf{E}_{m-1}^{drive},\mathbf{E}_m^{drive},\mathbf{E}_{m+1}^{drive},...)$. The operator $\mathbf{M}(\omega)$ has the form $\mathbf{M}(\omega) = \alpha^{-1}(\omega)\mathbf{I} - \mathbf{G}(\omega)$, where $\mathbf{I}$ is an identity operator and $\mathbf{G}(\omega)$ contains only geometrical information. We define the product between two vectors $|\mathbf{a}\rangle \doteq (...,\mathbf{a}_{m-1},\mathbf{a}_m,\mathbf{a}_{m+1},...)$ and $|\mathbf{b}\rangle \doteq (...,\mathbf{b}_{m-1},\mathbf{b}_m,\mathbf{b}_{m+1},...)$ as $\langle\mathbf{a}|\mathbf{b}\rangle = \sum_{m=-\infty}^{\infty}\bar{\mathbf{a}}_m\cdot\mathbf{b}_m$, where $\bar{\mathbf{a}}_m$ denotes the complex conjugate of $\mathbf{a}_m$. We note that the product defined here is different from that in Ref. 19. The formal solution to Eq. (3) is

$$|\mathbf{p}\rangle = \sum_\sigma \int_{k=-\pi/a}^{\pi/a} dk \frac{1}{\lambda_k^\sigma}|\hat{\mathbf{p}}_k^\sigma\rangle\langle\hat{\mathbf{p}}_k^\sigma|\mathbf{E}^{drive}\rangle, \quad (4)$$

where $\lambda_k^\sigma = \alpha^{-1}(\omega) - \kappa_k^\sigma$, $\kappa_k^\sigma$ is the eigenvalue of Green's operator $\mathbf{G}(\omega)$, and the corresponding "normalized" eigenvector is[24] $|\hat{\mathbf{p}}_k^\sigma\rangle \doteq \sqrt{a/(2\pi)}(...,\phi_{m-1}\hat{\mathbf{x}}_\sigma, \phi_m\hat{\mathbf{x}}_\sigma, \phi_{m+1}\hat{\mathbf{x}}_\sigma,...)$ with $\phi_m = e^{ikma}$ and $k$ being the wavenumber of the mode. The polarization index $\sigma$, which takes the values of 1, 2, or 3, indicates the $x$-, $y$-, or $z$-polarizations, respectively, with unit vectors $\hat{\mathbf{x}}_1$, $\hat{\mathbf{x}}_2$, and $\hat{\mathbf{x}}_3$. For $\sigma = 1$ and 2 (T mode), the dipole moments are perpendicular to $\hat{\mathbf{z}}$, while that for $\sigma = 3$ (L mode) is parallel to $\hat{\mathbf{z}}$.

[*]Electronic mail: phchan@ust.hk

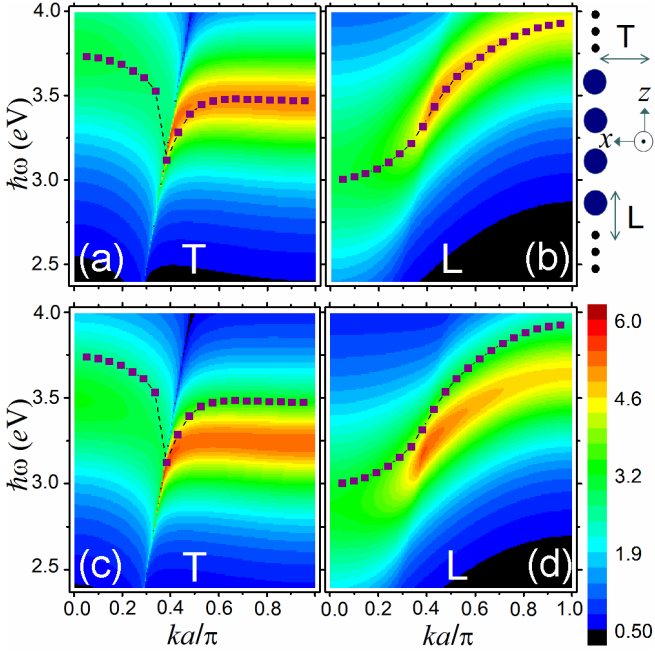
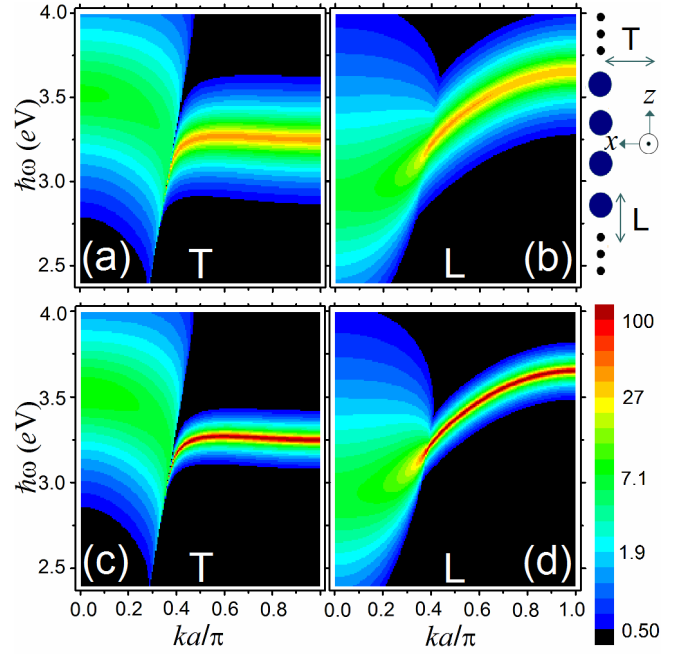

FIG. 1. (Color online) Intensity plot of $\mathrm{Im}(\alpha_k^\sigma(\omega))/d^3$ for the dispersion relation of a MNP chain with $\gamma = 0.7\ eV$. The symbol "T" ("L") represents the T (L) mode. (a) and (b) Quasi-static DA with radiation correction. (c) and (d) Dynamic DA. Squares connected by dashed lines are calculated by the root searching method using quasistatic DA with radiation correction.

FIG. 2. (Color online) Intensity plot of $\mathrm{Im}(\alpha_k^\sigma(\omega))/d^3$ calculated by dynamic DA. Symbol "T" ("L") represents the T (L) mode. (a) and (b) $\gamma = 0.1\ eV$. (c) and (d) $\gamma = 0.02\ eV$.

We note that the inverse of the eigenvalues $\alpha_k^\sigma(\omega) = [\lambda_k^\sigma(\omega)]^{-1}$ have the dimension of polarizability, and these effective "eign-mode polarizabilities" have the same physical meaning as single particle dipole polarizability $\alpha(\omega)$ except that it represent the collective characteristic response of the whole chain. As a spectral function, the peaks of $\mathrm{Im}(\alpha_k^\sigma(\omega))$[25] represent resonance modes, and in the limit of no dissipation, they define the band disperion. Therefore, the dispersion relation can be easily found by directly making an intensity (or contour) plot of $\mathrm{Im}(\alpha_k^\sigma(\omega))$ on the real $\omega-k$ plane for each polarization $\sigma$. The results for $\gamma = 0.7\ eV$ (chosen to compare with Ref. 12), which represents a lossy metal such as gold, are shown in Figure 1. In Fig. 1(a) and 1(b), the quasistatic DA with radiation correction is used. For comparison, the results of a finite MNP chain calculated by the root seaching method (Ref. 12) are also shown in the figures. For both T and L mode, the intensity plot of $\mathrm{Im}(\alpha_k^\sigma(\omega))$ agrees well with the real part of the complex root $\tilde{\omega}(k)$. In additional to the dispersion relation, the plots also show automatically the clear boundary of the light cone, especially for T mode. $\mathrm{Im}(\alpha_k^\sigma(\omega))$ is proportional to the extinction of the driving field and the width of an extinction peak is proportional to the mode quality. A wide width of the function $\mathrm{Im}(\alpha_k^\sigma(\omega))$ implies a lossy mode. Since $\mathrm{Im}(\alpha_k^\sigma(\omega))$ is a sum of absorption and radiation losses, both mechanism can broaden $\mathrm{Im}(\alpha_k^\sigma(\omega))$. Inside the light cone ($k < \omega/c$), the mode is not well-defined due to severe radiation loss. Outside the light cone, the spectral function has a narrower and conspicuous peaks. Thus, we see that such method offers a quantitative description and a qualitative understanding of guided modes at the same time.

We then consider the more rigorous but more complicated dynamic DA (Eq. (2)). The advantage of the method is that $\kappa_k^\sigma$ is already calculated, no additional effort is needed for the calculation of $\alpha_k^\sigma(\omega)$. As long as the single particle polarizability $\alpha(\omega)$ is given, the additional work is just a simple arithmetic calculation. We note that if we apply complex root searching for the dynamic DA, the task will be more tedious. The results by our approach are shown in Fig. 1(c) and 1(d). Again, the results obtained by complex root searching with quasistatic DA are compared. In addition to a small improvement in mode quality, we found that the dispersion curves are red-shifted by about $\Delta\hbar\omega \approx 0.25\ eV$ (about 25 nm in wavelength) when dynamic DA is employed. Since such correction to the mode frequency is not negligible, it shows that dynamic DA should be used in order to make comparison with experiments.

The above results are calculated using a high absorption rate ($\gamma = 0.7\ eV$), and Fig. 2 shows the results for $\gamma = 0.1\ eV$ and $0.02\ eV$ which are more relevant for silver. We keep the same $\omega_p$ so that a direct comparison between different absorption rates can be made. The qualities of the guided modes have a great improvement when abosrption rate is reduced, especially for the modes outside light cone where the finite line width originates purely from the resistive loss.

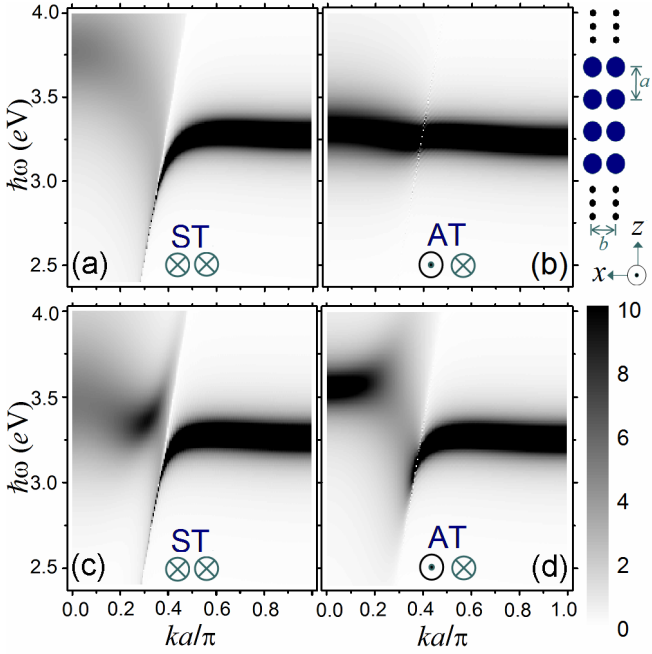

FIG. 3. (Color online) Intensity plot of $\text{Im}(\alpha_k^\sigma(\omega))/d^3$ for a parallel pair of MNP chains with $\gamma = 0.1$ eV. "ST" ("AT") represents the symmetric (anti-symmetric) T mode. (a) and (b) $b = a = 75$ nm. (c) and (d) $b = 5a = 375$ nm.

The ED method can also be applied to study the plasmonic modes of more complex systems such as chain arrays, and higher dimensional structures. As an example, we consider two parallel chains with the distance between chain axes being $b$. For simplicilty, we just consider the T modes with dipole moments perpendicular to the plane containing the two chains (as shown in Fig. 3). In this case, there are two T modes. The symmetric (antisymmetric) T mode, denoted by ST (AT), has each dipole of one chain being in-phase (out of phase) with the nearesest dipole of the other chain. Results for $b = a = 75$ nm are shown in Fig. 3(a) and 3(b). The ST mode has a dispersion relation similar to the T mode of a single chain, except a small increase in the slope inside light cone. The AT mode is very different. The whole band has a small but negative gradient, and the mode is well-defined even inside the light cone. Interesting results are also shown in Fig. 3(c) and 3(d) for the inter-chain distance $b = 5a = 375$ nm, which is close to the wavelength of light in free space. We can see a strong modification due to resonant coupling. Inside the light cone, the ST (AT) mode has a red (blue) shift for small $k$. Near the light line, the mode quality and the dispersion relation has obvious changes due to multiple scattering between chains, i.e. the plasmonic guided modes are coupled with quasi-modes guided in the space between the chains. Such inter-chain couplings can serve to manipulate the guided modes, and should warrant further studies.

In summary, we demonstrated that the ED method is very good for making a quick and accurate analysis of the plasmonic modes of MNP chains. Such method separates the material and geometrical properties, so its efficiency does not depend on the complexity of the material polarizability, and more accurate dynamic DA can be used without computation difficulty.

This work is supported by the Hong Kong RGC grant 600403. We also thank Lei Zhou for helpful discussions.